\documentclass[aps,twocolumn]{revtex4}
\usepackage{dcolumn}
\usepackage{graphicx}
\usepackage{graphicx,amssymb,amsmath}
\usepackage{natbib}
\begin{document}

%%%%%%%%%%%%%%%%%%%%%%%%%%%%%%%%%
%   The is version 9
%  Last changed on June 12, 2003 by AJL
%%%%%%%%%%%%%%%%%%%%%%%%%%%%%%%%%

\title{Non--universality of elastic exponents in random
bond--bending networks}

\author{D.A. Head$^{1,2}$}%\email[]{dhead@nat.vu.nl}

\author{F.C. MacKintosh$^{1,2}$}%\email[]{fcm@nat.vu.nl}

\author{A.J. Levine$^{2,3}$}%\email[]{levine@physics.umass.edu}

\affiliation{$^{1}$Division of Physics \& Astronomy, Vrije Universiteit 
1081
  HV Amsterdam, The Netherlands}
\affiliation{$^{2}$The Kavli Institute for Theoretical
  Physics, University of California, Santa Barbara CA 93106, USA}
\affiliation{$^{3}$Department of Physics, University of Massachusetts,
Amherst MA 01060, USA}

\date{\today}

\begin{abstract}
We numerically investigate the rigidity percolation transition in two--dimensional flexible, random rod networks with freely rotating cross-links.
Near the transition, networks are dominated by bending modes
and the elastic modulii vanish with an exponent
exponent $f=3.0\pm0.2$,
in contrast with central force percolation which shares the
same geometric exponents.
This indicates that universality for geometric quantities
does not imply universality for elastic ones. The implications of this result for actin-fiber networks is discussed.
\end{abstract}

\pacs{PACS numbers: 64.60.Ak, 62.20.Dc, 46.25.-y}

\maketitle

In contrast with ordered crystals, disordered materials typically 
exhibit a complex connection between their collective mechanical 
properties and the underlying interactions between their constituent 
elements~\cite{Rubenstein:03}. In this Rapid Communication, we study 
this connection between effective elastic theories of a flexible rod 
network in two dimensions and the fundamental elastic properties of 
individual rods, along with statistical measures of the network 
connectivity.  We focus on cross-link densities to explore the 
break-down of collective rigidity of the network. This vanishing of the 
static shear modulus in related 
systems~\cite{review,site,pebble,random_rigidity, 2D_BB,3D_BB} has been 
termed the rigidity percolation transition in analogy to the better 
understood scalar, or conductivity, percolation 
transition~\cite{conductivity}. Previously~\cite{Head:03} we 
investigated the internal deformation field of a 2d flexible rod network 
away from the critical point of rigidity percolation, and found a 
cross-over from affine to nonaffine deformation upon changing the mean 
density between cross links in the network and/or the inherent 
flexibility of individual rods, which wasl also independently 
corroborated by another group~\cite{Wilhem:03}.

The rigidity percolation transition has been previously studied numerically in a number of model, disordered systems with variations in both the nature of the links (bonds) and the disorder of the network. While the rigidity transition is first order on the Bethe lattice~\cite{Duxbury:95} and in random networks of infinitely rigid rods~\cite{Obukhov:95}, in all other random network models it is continuous and characterized by a diverging length scale over which the material acquires a finite, static shear modulus. From the outset, it is important to distinguish between two different classes of physical observables when discussing the long length scale physics at the transition:(i) {\it geometric\/} quantities which describe the fractal structure of the percolating rigid cluster at the transition; and (ii) {\it elastic\/} properties of the material near the percolation point.  In the first category are the exponents $\nu$ and $\beta$, which describe the divergence of the correlation length as one approaches the transition and the probability that a link will be part of the percolating cluster, respectively. In the second category is $f$, which describes the power law dependence of the system's elastic moduli upon the approach to the percolation transition from the rigid phase (see Eq.~\ref{e:fit1}).

In lattice models, the control parameter, as for scalar percolation,
is the probability of the presence of a link in the diluted lattice. 
It appears that the nature of disorder is a relevant variable, since for purely central forces between network nodes, site and bond disorder exhibit quantitatively different scaling regimes near the critical point~\cite{review, site}. In addition, a third universality class has been postulated in the so-called ``bond--bending'' model, where bending lattice edges and rotation at vertices cost energy~\cite{2D_BB,3D_BB,astrom}. This suggests that the introduction of a bending modulus in the model is a relevant perturbation at the transition.

Pertinent to the current work are the simulations of Latva-Kokko {\it et al.\/}~\cite{pebble,random_rigidity}.  These investigations, departing from previous work, turned to off--lattice simulations. They constructed two-dimensional random rod networks and applied a topological approach to investigate the rigidity percolation point in random rod networks with Hookean central forces and a bending modulus. The introduction of a bending modulus is vital at this stage since the random spring network that is not prestressed has zero--frequency deformation modes at any finite density of links, and is thus always nonrigid~\cite{G_Zero}.  Latva-Kokko {\it et al.\/} considered two variants of their model of flexible rods distinguished by the constraint forces imposed at a cross link: one with cross links that fix the angle between intersecting rods by applying local constraint torques, and freely rotating bonds at cross links in the other. In both cases they found that the {\it geometric\/} exponents are consistent with those of the rigidity percolation transition in a diluted lattice with central forces. This result suggests that the introduction of bending forces are not relevant perturbations at the critical point in the following restricted sense: the exponents associated with length scale and geometry of the spanning, rigid cluster appear to be universal. We use this result later. Their approach, however did not allow them to investigate the scaling exponent for the elastic quantities of their model. 

This Rapid Communication presents a study of the elastic properties of the rod network identical to the model system studied by Latva-Kokko {\it et al.\/}. We, however, concentrate on the mechanical aspects of the network near the transition. It is important to note that this work relates to models of semi-flexible polymer networks with freely rotating bonds~\cite{wlc}. This detail of the model distinguishes it from its other close antecedent, the bond-bending model of S.\ Feng {\it et al.\/}.
That lattice--based model has been particularly well described by both numerical simulation and real-space renormalization group techniques~\cite{Feng:85}.
Our model differs in that the two filaments that cross
at each node contribute independently to that node's
bending energy; there is no energy cost for relative
rotation between rods.
Previous lattice calculations suggest that the scaling of the elastic constants near the transition depend on such details of the network; with this in mind, we seek herein to explore the rigidity collapse of sparse actin networks.

The principal result of this communication is that the mechanical properties of the flexible rod network at the rigidity percolation transition are distinct from previously investigated models of either lattice--based bond-bending networks or central force networks. This is true despite the fact that the scaling of the size of the percolating cluster and its fractal geometry (as determined by Latva-Kokko {\it et al.\/}~\cite{pebble}) are identical to that of central force networks. This point highlights the physical independence of geometric/topological exponents and the elastic exponents of the network.
We propose that whilst some degree of universality in the
elastic properties of disordered systems may exist,
such universality classes are smaller and more numerous than
for geometric properties;
that is, models with the same geometric exponents $\nu$
and $\beta$ may have distinct elastic exponents~$f$.
Our model system below provides a concrete
example of this claim.
We suspect that the underlying cause of this non--universality is that while the rigidity transition itself depends on large scale topological properties of the network that are insensitive to the local details of stress transmission, the moduli of the rigid network near the transition are sensitively determined by only a vanishingly small set of paths for stress propagation.  The moduli thus depend on local details of how stress is transmitted through the small set network nodes along the weak points of this path. Different force laws at {\it e.g.\/} cross links can therefore have profound effects on the way in which the moduli vanish at the transition.

We numerically evaluated the elastic moduli of random networks of rods at $T=0$,
with an energy $\delta{\mathcal H}$ per unit length $\delta s$ given by
\begin{equation}
\frac{\delta{\cal H}}{\delta s}
=
\frac{\mu}{2}
\left(
\frac{\delta l}{\delta s}
\right)^{2}
+
\frac{\kappa}{2}
\left(
\frac{\delta\theta}{\delta s}
\right)^{2}
\label{e:hamiltonian}
\end{equation}
The first term on the right hand side describes an elastic restoring force for changes in relative length of the rods, $\delta l/\delta s$, with a spring modulus~$\mu$. Bending by an angle $\delta\theta$ incurs an energy cost given by the second term, where the bending modulus $\kappa$ is the same as in the wormlike chain model~\cite{wlc}. Cross linked rods are coupled by imposing the same coordinates at intersections.

\begin{figure}[htpb]
  \centering
  \includegraphics[width=8cm]{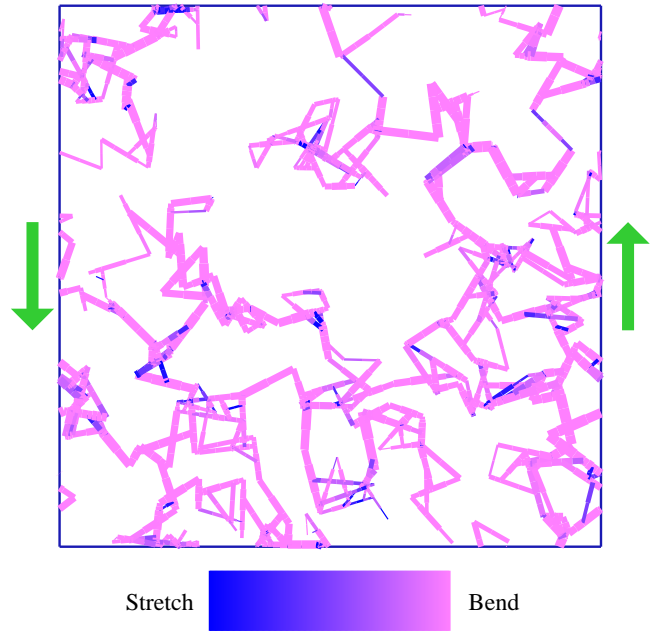}
  \caption{{\em (Color online)}
The energy density for a network at the
transition $L/l_{\rm c}\approx5.933$ under an applied shear strain.
The line thickness is proportional to the logarithm of the energy density per unit length.
Apparent stressed `dangling' ends are artefacts of
numerical noise and make no significant difference to
the measured quantities discussed below.
The calibration bar shows what proportion of the energy is due to stretching.
For clarity, a small $W=7\frac{1}{2}L$ shear cell is shown.}
\label{f:eg_enden}
\end{figure}

The simulation method is described in detail elsewhere~\cite{inprep} (where we also discuss behavior away from the transition), but in brief: rods of length $L$ are deposited with random orientation and position into a two--dimensional shear cell of dimensions $W\times W$. Each intersection is identified as a cross link, the mean distance between which (as measured along a rod) is denoted $l_{\rm c}$\,, so that the mean number of crosslinks per rod is $L/l_{\rm c}-1$.
Deposition continues until the required cross linking density $L/l_{\rm c}$ has been reached.
The system Hamiltonian is constructed from (\ref{e:hamiltonian}), and the mechanical equilibrium  configuration found by the preconditioned conjugate gradient method under the constraint of an applied shear or uniaxial strain. An example is given in Fig.~\ref{f:eg_enden}.

\begin{figure}[htpb]
\centering
  \includegraphics[width=8.5cm]{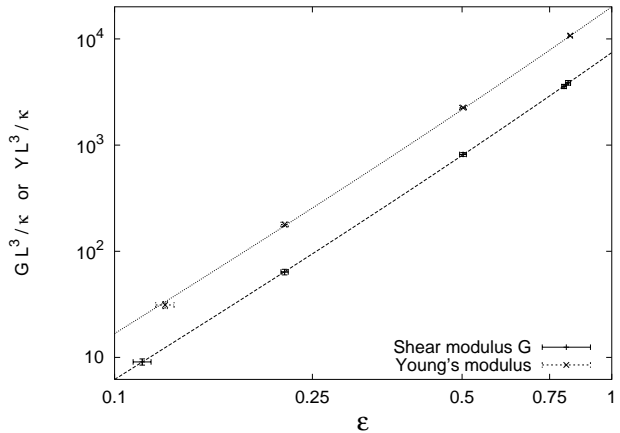}
\caption{Dimensionless shear modulus (lower data set)
and Young's modulus (upper set) for
$l_{\rm b}/L=0.006$ versus $\varepsilon$
on log--log axes.
The best--fit of (\ref{e:fit1}) to the data for $G$ is plotted,
and then shifted vertically to show agreement with $Y$.
}
\label{f:Y_near_cp}
\end{figure}

There are four lengths in the problem: the system size and rod length $W$, $L$ respectively;
the mean distance between cross links $l_{\rm c}$ and a length characterizing the flexibility of the rods, $l_{\rm b} =\sqrt{\kappa / \mu}$. We take $L/l_{\rm c}$ as our dimensionless measure of density rather than $q=NL^{2}$ with $N$ the number of rods per unit area, as in~\cite{pebble,random_rigidity,G_Zero}. It is straightforward to convert between the two measures using the expression derived in the Appendix. Rigidity percolation occurs at $(L/l_{\rm c})_{\rm trans}\approx5.933$~\cite{pebble}.  For $L/l_{\rm c}$ slightly above this critical point, where $\varepsilon=(L/l_{\rm c})/(L/l_{\rm c})_{\rm trans}-1 \sim 0^+$, we write the shear modulus $G$ as
\begin{equation}
G= \frac{\kappa}{L^{3}} f\left( \varepsilon, \frac{l_{\rm b}}{L}, \frac{W}{L} \right)
\end{equation}
with a similar expression for the Young's Modulus, $Y$. For $\varepsilon>0$, and sufficiently large system sizes $W$, both $G$ and $Y$ depend only on $\varepsilon$ and $l_{\rm b}/L$. Fig.~\ref{f:Y_near_cp} shows $G$ and $Y$ versus $\varepsilon$ for systems where this large--$W$ limit has been reached.
Both moduli vanish continuously as $\varepsilon\rightarrow0^{+}$,
confirming both the nature and location of the transition as claimed in~\cite{pebble}, to within our data resolution~\cite{further-conf}.

\begin{figure}[htpb]
  \centering
  \includegraphics[width=8.5cm]{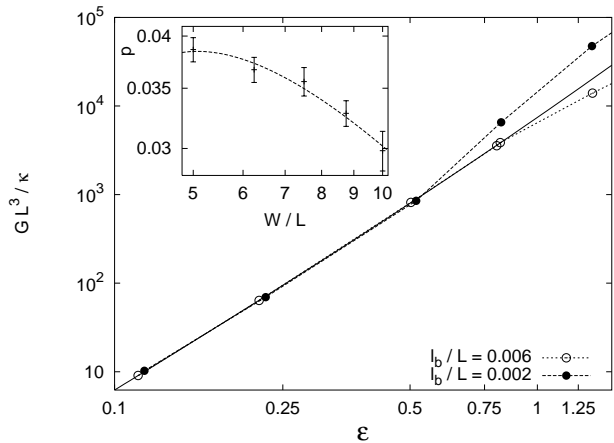}
  \caption{The dimensionless shear modulus vs.\ $\varepsilon$ for two different $l_{\rm b}/L$ on log--log axes. For clarity the error bars are not plotted in this figure (see Fig.~\ref{f:Y_near_cp}), but are no larger than the symbols.
The solid line is a best fit to (\ref{e:fit1}) for $l_{\rm b}/L=0.006$ and $\varepsilon<1$,
with each point weighted by its fractional deviation from the fit.
{\em (Inset)}~The fraction of stretching energy to the total,
$p$, as a function of system size for $l_{\rm b}/L=0.006$.
The line is a best--fit to the scaling form in~\cite{3D_BB}.}
\label{f:G_near_cp}
\end{figure}

Fig.~\ref{f:G_near_cp} shows that $GL^{3}/\kappa$ is independent of $l_{\rm b}/L$ for sufficiently small~$\varepsilon$,
{\em i.e.} $G$ is {\em independent of $\mu$ near the transition}.
This suggests that the transition is dominated
by bending modes, and we can infer that $G$ would vanish if there were only central force terms in agreement with \cite{G_Zero}.
We can confirm the dominance of bending modes at the
transition by measuring the proportion of stretching energy
for increasing $W$.
As shown in inset of this figure, %Fig.~\ref{f:G_near_cp},
this fraction is always small, $<$4\%,
and may vanish as $W\rightarrow\infty$, as suggested by the figure and in agreement with
both~\cite{G_Zero} and
our observation that $G\propto\kappa$.

By fitting the data to the functional form
\begin{equation}
G=A\varepsilon^{f}(1+B\varepsilon)
\label{e:fit1}
\end{equation}
we find $f=3.0\pm0.2$, consistent with~\cite{Wilhem:03}, where
the error estimate gives the spread of values when the fitting procedure is repeated over different data subsets. $B$ is always ${\mathcal O}(1)$, suggesting
(\ref{e:fit1}) is a `sensible' choice.
This is compared to known values in Table~\ref{table}.
%in 2D are~\cite{review,2D_BB,forces,site}: $1.35\pm0.06$ for site percolation, $3.57\pm0.3$ for central--force bond percolation, and $3.96\pm0.04$ for the bond--bending model.
The only possible equality is with central force bond percolation;
however, since the modulus near the transition is dominated by {\em bending} forces, we do not believe that this near agreement is meaningful.
%Note that the $f$ found for the only other bond--bending
% model differs by  $\approx4$ error bars, indicating a lack
%  of universality of the bond-bending model moduli at the transition.
%%%% BUT ..... the BB model has different geometric exponents,
%%%% so this doesn't support our case.
The fit for $G$ can be made to agree with the
Young's modulus data by simply rescaling $A\rightarrow A^\prime$ with $f$ and
$B$ fixed, as shown in Fig.~\ref{f:Y_near_cp} for $l_{\rm b}/L=0.006$.
$A^\prime/A=2.7\pm0.2$, so that the Poisson ratio $\sigma\equiv Y/2G-1=0.35\pm0.1$ at the transition, consistent with the claimed `universal' value of $\frac{1}{3}$~\cite{poisson}.

\begin{table}
\caption{\label{table}Quoted $f$ for 2d uncorrelated networks}
\begin{ruledtabular}
\begin{tabular}{ccc}
Model & $f$ & Ref. \\
\hline\\
Site percolation & $1.35\pm0.06$ & \cite{review,site}\\
Central force percolation & $3.57\pm0.3$ & \cite{review,site} \\
Bond--bending model & $3.96\pm0.04$ & \cite{review,2D_BB,forces} \\
This work & $3.0\pm0.2$ & n/a
\end{tabular}
\end{ruledtabular}
\end{table}

%\begin{figure}
%\centering
% \includegraphics[width=4.0cm,angle=270]{finsize.ps}
%\caption{The dimensionless shear modulus $GL^{3}/\kappa$
%versus the dimensionless system size $W/L$
%at the transition $L/l_{\rm c}\approx5.933$ on log--log axes,
%with $l_{\rm b}/L=0.006$.
%The two lines are best--fits to expression (\ref{e:fit2})
%for the two given exponents $f/\nu$.
%}
%\label{f:G_with_W}
%\end{figure}

Independent confirmation of $f$ can be found by
placing the system at the critical point $\varepsilon=0$
and considering the variation of $G$ with $W$~\cite{review,2D_BB,3D_BB}.
Scaling arguments suggest that $G\sim W^{-f/\nu}$ for sufficiently large~$W$, where $\nu$ is the exponent describing the divergence of the correlation length, which is known for these networks to be in the range of $\nu=1.17\pm0.02$~\cite{pebble}. We attempted to fit our data to the same functional form used in~\cite{3D_BB}, %which includes a choice of leading analytical and non--analytical %corrections to scaling,
%\begin{equation}
%G=CW^{-f/\nu}[1+D/W+E/\ln(W)]
%\label{e:fit2}
%\end{equation}
%Unfortunately, given
but the precision with which we can extract the exponent using this approach is not sufficient to distinguish our results from either central force or bond--bending models.
%This is clearly demonstrated in Fig.~\ref{f:G_with_W},
%where we see that $f/\nu$ can be equally well fit using
% Eq.~(\ref{e:fit2}),
%$f/\nu=1.97$, or $2.97$ by adjusting the constants, $C$, $D$ 
% and $E$ appropriately. 
The number of networks for each attempted $W$ ranged from
400 to 2000, comparable to lattice models which give
much smaller errors.
Our lack of precision is likely due to the random nature of our networks, in which the distance between crosslinks varies continuously down to zero so that the strength of coupling
between connected nodes can vary greatly, introducing
a significant additional noise source into the problem.

In summary, we have shown that the elastic moduli of flexible random networks with freely rotating crosslinks scale as $\varepsilon^{f}$ near to
rigidity percolation. The modulus itself is dominated by bending modes near the transition and vanishes with an exponent $f=3.0\pm0.2$. This exponent differs from previous reported exponents for models that include bond bending.  Our introduction of bending forces, motivated as it was by modelling cross linked F-actin networks, differs in detail from previous work. Our results suggest that, in contrast to the geometric exponents previous reported for this system, the modulus exponent is more highly model dependent. To address the experimental implications of these results for physical actin networks, we note that the continuous transition with true scale invariant behavior exists at only zero temperature. However, zero temperature phase transitions can have experimental consequences at finite temperature~\cite{QCP} as long as it is possible to experimentally access the critical region. We expect that in sparse actin networks the algebraic decay of the static shear modulus will be cut off only by the entropic elasticity of the network~\cite{zero_T}, a small quantity for a stiff polymer network. The experimental observation of this predicted decay, a signature of the $T=0$ transition at finite temperature, may be possible, but such quantitative predictions await a more complete description of the critical regime.

{\em Acknowledgements:} AJL thanks A.D.\ Dinsmore for useful discussions and would like to acknowledge the hospitality of the Vrije
Universiteit. DAH was partly funded by a European Community Marie Curie Fellowship. 
This work is  supported in part by the National Science Foundation under Grant Nos.\ DMR98-70785 and PHY99-07949.

{\it Appendix:} Here we derive the relationship between $L/l_{\rm c}$ and the quantity $q=NL^{2}$,
where $N$ is the number of rods per unit area, as used in~\cite{pebble,random_rigidity,G_Zero,Wilhem:03}.
Suppose a rod of length $L$ lies along the $x$--axis of a $W\times W$ box,
where $W\gg L$ so that the box shape should not matter. The probability of a second rod, deposited with orientation $\theta$ to the $x$-axis and random position, intersecting the $x$--axis is $(L/W)\sin\theta$. Hence the probability of intersecting the first rod is $(L/W)^{2}\sin\theta$, which, when uniformly averaged over all $\theta\in[0,\pi]$,
becomes $p=2L^{2}/(\pi W^{2})$.

Since the rods are deposited at random, $p$ is independent of how many other crosslinks each rod has. The mean number of intersections per rod is therefore simply $p$ times the total number of rods in the system $NW^{2}$, {\em i.e.} $pNW^{2}=2L^{2}N/\pi=2q/\pi$
(assuming $NW^{2}\gg1$).
The distribution $P_{n}$ of the number $n$ of cross links for any given rod is Poisson with a mean $2q/\pi$.
%\begin{equation}
%P_{n}=
%\left(
%\frac{2q}{\pi}
%\right)^{n}
%\frac{{\rm e}^{-2q/\pi}}{n!}
%\quad.
%\label{e:P_n}
%\end{equation}
If each intersection is imagined to `cut' the rod into $n+1$ line segments, then the mean length of all such segments is just $L/(\langle n\rangle+1)=L/(2q/\pi+1)$. However, $l_{\rm c}$ is defined between crosslinks, {\em not} between crosslinks and the ends of the rods
(which are dangling and thus do not contribute to the stress). In fact, only $n-1$ segments  contribute to $l_{\rm c}$\,. Thus to evaluate $l_{\rm c}$, we must average the quantity $L/(n+1)$ over all valid $P_{n}$ with a weighting $n-1$,
and normalize accordingly, {\em i.e.}
\begin{equation}
l_{\rm c}
=
\frac{\sum_{n=2}^{\infty}\frac{L}{n+1}(n-1)P_{n}}
{\sum_{n=2}^{\infty}(n-1)P_{n}}
\end{equation}
Using the shorthand $\lambda=2q/\pi$ and the standard properties of the Poisson distribution, we find:
\begin{equation}
\frac{L}{l_{\rm c}}
=
\frac{\lambda+{\rm e}^{-\lambda}-1}
{1+{\rm e}^{-\lambda}+\frac{2}{\lambda}
({\rm e}^{-\lambda}-1)},
\label{e:final}
\end{equation}
which is a monotonic increasing function of $\lambda$.
Note that $L/l_{\rm c}\sim 2q/\pi$ as $q\rightarrow\infty$,
and also that $L/l_{\rm c}\rightarrow3$ as $q\rightarrow0^{+}$, as it should since (in this limit) the dominant contributions to $l_{\rm c}$ will come from rods with $n=2$ crosslinks, for which $l_{\rm c}$ is indeed $L/3$.


\begin{thebibliography}{LL}
\bibitem{further-conf}
To further confirm the continuous nature of the transition note that
smaller $\varepsilon$ requires a larger $W$ for convergence, which is
consistent with a diverging correlation length.
\end{thebibliography}
\end{document}